# The ferromagnetic order in graphene

L. Wojtczak

*Solid State Physics Department, University of Lodz, ul. Pomorska 149/153, 90236 Lodz, Poland*

The conditions for spontaneous magnetization in a single graphene sheet are discussed in the context of Mermin-Wagner theorem. It is rigorously proved that at any nonzero temperature the graphene monolayer is nonmagnetic as long as its intrinsic symmetry is preserved. Any electronic fluctuations breaking the pseudospin conservation law can be responsible for the existence of ferromagnetic order inside the graphene structure.

As early as in the thirties, Peierls [1] and Landau [2] argued that systems whose structure is periodic in one or two dimensions cannot exist. This is due to the important role of large long-wavelength fluctuations which, in some sense, wash out the assumed long-range periodicity. Much later Hohenberg [3] discussed a respective theorem for one- and two-dimensional superfluids and superconductors. In the case of the 2D lattice the fluctuations are such that each atom performs a motion around its assumed equilibrium position whose extent is much larger than the average periodicity length and is in fact weakly divergent in the thermodynamic limit. However, these large motions are still performed around well-defined equilibrium positions and it turns out that macroscopically large parts of the system move almost in unison. Alternatively, it follows that macroscopically large finite systems can show an effective long-range ordering. The 2D harmonic lattice [4] is the simplest Landau-Peierls 2D system which, although not having strict long-range ordering due to large fluctuations, displays a slow decay of the appropriate correlation functions. For example, the displacement correlation function decays weakly enough as function of distance, so that the appropriate susceptibilities may diverge [4]. This slow decay results in the observable properties which differ only slightly from those of strictly ordered systems. These slight differences are well defined and uniquely characterize the quasi-long-range order of such 2D systems. This result, which seems surprising at first, follows from the fact that the large fluctuations occur over long distances, while the short-range fluctuations which determine the validity of the harmonic approximation are not anomalously large in 2D. Similarly, Bloch had shown earlier [5] that ferromagnetism does not exist in two-dimensional (2D) Heisenberg systems. Furthermore, using an inequality due to Bogoliubov [6], Mermin and Wagner [7] proves rigorously the absence of both ferromagnetic and antiferromagnetic order in one- (1D) and two-dimensional spin systems described by the Heisenberg Hamiltonian. This theorem concerning the absence of magnetic ordering in 1D and 2D can be proved not only for systems described by the Heisenberg Hamiltonian, but for many others as well [8-10]. In this context, one knows quite generally that strict long-range order does not exist at and below two dimensions for systems where the magnetic order parameter has a continuous symmetry.

In 2004, the group of K. S. Novoselov and A. K. Geim [11], provided the first and unexpected proof for the existence of true (free-standing) 2D crystals. Previously, it was assumed that graphene cannot exist in the flat state and should scroll into nanotubes to decrease the surface energy. With the experimental isolation of graphene [11], the hopping model has been shown to give an effective description of these real materials. This theory was first developed by Philip R. Wallace in 1947 [12] as an approximation trying to understand the electronic properties of more complex, 3 dimensional graphite. He did not use the word graphene and referred to "a single hexagonal layer".

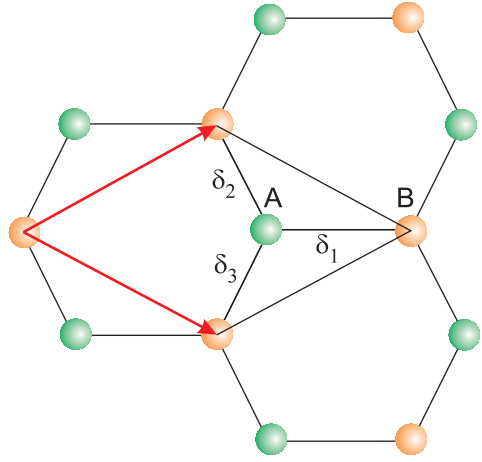

FIG. 1: Lattice structure of graphene constructed by two interpenetrating triangular sublattices.

The carbon atoms in graphene form 2D honeycomb network with two nonequivalent atomic sites per unit cell (Fig. 1). The symmetry between sublattices formed by these two atomic sites is responsible for massless character of Dirac fermions and the lack of the bandgap in a single graphene sheet. Its cone-shaped valence and conduction bands touch each other at the corners of the Brillouin zone exhibiting the linear dispersion relation at their vicinity.

In the following we shall discuss the Mermin-Wagner theorem using a single band model for the $\pi$ electrons of graphene described by the Hubard Hamiltonian:

$$H = t \sum_{\langle i,j \rangle, \sigma=\uparrow,\downarrow} \left( a^+_{\sigma,i} b_{\sigma,j} + b^+_{\sigma,j} a_{\sigma,i} \right) + U \sum_{i=A,B} n_{i\sigma} n_{i-\sigma} + \frac{1}{2} h \sum_{i=A,B} \left( n_{i\uparrow} - n_{i\downarrow} \right) e^{-i\vec{q}\vec{R}_i} \quad (1)$$

where $n_{A\sigma} = a^+_\sigma a_\sigma$ and $n_{B\sigma} = b^+_\sigma b_\sigma$. The operators $a^+_{\sigma,i} (a_{\sigma,i})$ and $b^+_{\sigma,j} (b_{\sigma,j})$ create (annihilate) an electron of spin $\sigma$ at location $I$ (= $i$ or $j$) on the A and B sublattices, respectively. These fermion operators obey the canonical anticommutation relations: $\{c_{j\sigma}, c^+_{k\sigma'}\} = \delta_{jk} \delta_{\sigma\sigma'}$ and $\{c_{j\sigma}, c_{k\sigma'}\} = \{c^+_{j\sigma}, c^+_{k\sigma'}\} = 0$. The first term of Hamiltonian $H$, involving hopping parameter $t$, takes into account nearest-neighbor interactions. $U$ is the energy associated with having two electrons at the same lattice site while the last term corresponds to a space-dependent external magnetic field.

The quantity we wish to compute is the magnetization induced by the magnetic field $h$. If the magnetic field can induce a magnetization that tends to a constant non-zero value when $h$ tends to zero, the system will exhibit a spontaneous magnetization. If the induced magnetization tends to zero as *h* tends to zero, there will be no spontaneous magnetization. Of course, the case $\vec{q} = 0$ corresponds to ferromagnetism while the interaction for arbitrary $\vec{q}$ probes more complicated kinds of ordering.

Following Mermin and Wagner [5], we make use of the Bogoliubov inequality:

$$\frac{1}{2}\langle\{A, A^+\}\rangle|\langle[[C,H],C^+]\rangle| \geq k_B T |\langle[C,A]\rangle|^2 \qquad (2)$$

For the operators *C* and *A* we choose:

$$C = S_+(\vec{p}),\ A = S_-(-\vec{p}-\vec{q}) \qquad (3)$$

and we define the spin operators for each sublattice in the usual way, i.e.:

$$S_+(R_{i=A,B}) = c^+_{i\uparrow}c_{i\downarrow};\ S_-(R_{i=A,B}) = c^+_{i\downarrow}c_{i\uparrow};\ S_z(R_{i=A,B}) = \frac{1}{2}(n_{i\uparrow} - n_{i\uparrow}) \qquad (4)$$

We also introduce the spatial Fourier transforms:

$$S_\alpha(\vec{p}) = \sum_{R_{i=A,B}} \exp(i\vec{p}\vec{R}_i) S_\alpha(\vec{R}_i) \qquad (5)$$

of the spin operators, where $S_\alpha$ can be $S_+, S_-, S_z$.

The evaluation of the Bogoliubov inequality can be conducted in two steps. In the first step we calculate the Hamiltonian independent quantities:

$$\{A, A^+\} = \{S_-(-\vec{p}-\vec{q}), S_+(\vec{p}+\vec{q})\} =$$
$$\sum_{\vec{R}_A, \vec{R}_B}\left[e^{i(\vec{p}+\vec{q})(\vec{R}_A-\vec{R}_B)}\left(a^+_\uparrow a_\downarrow b^+_\downarrow b_\uparrow + b^+_\downarrow b_\uparrow a^+_\uparrow a_\downarrow\right) + e^{i(\vec{p}+\vec{q})(\vec{R}_B-\vec{R}_A)}\left(b^+_\uparrow b_\downarrow a^+_\downarrow a_\uparrow + a^+_\downarrow a_\uparrow b^+_\uparrow b_\downarrow\right)\right] \qquad (6)$$
$$+ N_A \sum_{\vec{R}_A}\left[n_{A\downarrow} + n_{A\uparrow} - 2n_{A\downarrow}n_{A\uparrow}\right] + N_B \sum_{\vec{R}_B}\left[n_{B\downarrow} + n_{B\uparrow} - 2n_{B\downarrow}n_{B\uparrow}\right]$$

and

$$[C, A] = [S_+(\vec{p}), S_-(-\vec{p}-\vec{q})] = N_A S_z^A(-\vec{q}) + N_B S_z^B(-\vec{q})$$
$$+ \sum_{\vec{R}_A, \vec{R}_B}\left[e^{i\vec{p}(\vec{R}_A-\vec{R}_B)}e^{-i\vec{q}\vec{R}_B}\left(a^+_\uparrow a_\downarrow b^+_\downarrow b_\uparrow - b^+_\downarrow b_\uparrow a^+_\uparrow a_\downarrow\right) + e^{i\vec{p}(\vec{R}_B-\vec{R}_A)}e^{-i\vec{q}\vec{R}_A}\left(b^+_\uparrow b_\downarrow a^+_\downarrow a_\uparrow - a^+_\downarrow a_\uparrow b^+_\uparrow b_\downarrow\right)\right] \qquad (7)$$

while the second step contains the determination the Hamiltonian dependent quantity:

$$[[C,H],C^+] = [[S_+(\vec{p}), H], S_-(-\vec{p})] = t\sum_{\vec{\delta}}\left[\left(e^{-i\vec{p}\vec{\delta}} - 1\right)\left(a^+_\uparrow b_\uparrow + b^+_\downarrow a_\downarrow\right) + \left(e^{i\vec{p}\vec{\delta}} - 1\right)\left(a^+_\downarrow b_\downarrow + b^+_\uparrow a_\uparrow\right)\right]$$
$$+ h\left[N_A S_z^A(-\vec{q}) + N_B S_z^B(-\vec{q})\right] \qquad (8)$$

where we denoted by $\vec{\delta}$ the relation between sublattices $\vec{\delta} = \vec{R}_B - \vec{R}_A$ (see Fig. 1). Expanding the functions $\left(e^{\pm i\vec{p}\vec{\delta}} - 1\right)$ up to the first order while introducing the above results to the Bogoliubov inequality and summing both sides of it over $\vec{p}$, we may conclude that:

$$\frac{1}{2}\left(N_A \sum_{\vec{R}_A}[\langle n_{A\downarrow}\rangle + \langle n_{A\uparrow}\rangle - 2\langle n_{A\downarrow}n_{A\uparrow}\rangle] + N_B \sum_{\vec{R}_B}[\langle n_{B\downarrow}\rangle + \langle n_{B\uparrow}\rangle - 2\langle n_{B\downarrow}n_{B\uparrow}\rangle]\right) \geq$$

$$k_B T \left| N_A \langle S_z^A(-\vec{q})\rangle + N_B \langle S_z^B(-\vec{q})\rangle \right|^2 \tag{9}$$

$$\sum_{\vec{p}}\left[t\left|\sum_{\vec{\delta}}\vec{p}\vec{\delta}\left(\langle a_\downarrow^+ b_\downarrow\rangle + \langle b_\uparrow^+ a_\uparrow\rangle - \langle a_\uparrow^+ b_\uparrow\rangle - \langle b_\downarrow^+ a_\downarrow\rangle\right) + h\left(N_A \langle S_z^A(-\vec{q})\rangle + N_B \langle S_z^B(-\vec{q})\rangle\right)\right|\right]^{-1}$$

The quantities $N_A$ and $N_B$ denote the lattice sites of $A$ and $B$ sublattices, respectively. It is obvious that:

$$\sum_{\vec{R}_{i=A,B}}[\langle n_{i\downarrow}\rangle + \langle n_{i\uparrow}\rangle - 2\langle n_{i\downarrow}n_{i\uparrow}\rangle] \leq \sum_{\vec{R}_{i=A,B}}[\langle n_{i\downarrow}\rangle + \langle n_{i\uparrow}\rangle] \leq 2N_{i=A,B} \tag{10a}$$

and

$$\left|\sum_{\vec{\delta}}\vec{p}\vec{\delta}\left(\langle a_\downarrow^+ b_\downarrow\rangle + \langle b_\uparrow^+ a_\uparrow\rangle - \langle a_\uparrow^+ b_\uparrow\rangle - \langle b_\downarrow^+ a_\downarrow\rangle\right) + h\left(N_A \langle S_z^A(-\vec{q})\rangle + N_B \langle S_z^B(-\vec{q})\rangle\right)\right| \leq$$

$$\sum_{\vec{\delta}}\vec{p}\vec{\delta}\left|\langle a_\downarrow^+ b_\downarrow\rangle + \langle b_\uparrow^+ a_\uparrow\rangle - \langle a_\uparrow^+ b_\uparrow\rangle - \langle b_\downarrow^+ a_\downarrow\rangle\right| + h\left|N_A \langle S_z^A(-\vec{q})\rangle + N_B \langle S_z^B(-\vec{q})\rangle\right| \tag{10b}$$

and the Bogoliubov inequality can be rewritten in the following form:

$$\left|S^Z\right|^2 \leq \frac{N_A^2 + N_B^2}{k_B T}\left[\sum_{\vec{p}}\frac{1}{\Omega \vec{p} + h|S^Z|}\right]^{-1} \tag{11}$$

where

$$\Omega = \sum_{\vec{\delta}}\vec{\delta}\left|\langle a_\downarrow^+ b_\downarrow\rangle + \langle b_\uparrow^+ a_\uparrow\rangle - \langle a_\uparrow^+ b_\uparrow\rangle - \langle b_\downarrow^+ a_\downarrow\rangle\right| \tag{12a}$$

$$S^Z = N_A \langle S_z^A(-\vec{q})\rangle + N_B \langle S_z^B(-\vec{q})\rangle \tag{12b}$$

We now replace the sum over $\vec{p}$ by an integral. The inequality is strengthened if we integrate only over the first Brillouin zone, so if $p_0$ is the distance of the nearest Bragg plane from the origin in $\vec{p}$ space, then in two dimensions:

$$\left|S^Z\right|^2 \leq \frac{N_A^2 + N_B^2}{k_B T}\left[\frac{1}{p_0/\Omega + (h|S^Z|/\Omega^2)\ln(h|S^Z|/(\Omega p_0 + h|S^Z|))}\right] \tag{13}$$

It is clearly seen that as far as $\Omega$ is different from zero the magnetic order can be expected in the graphene monolayer even when $h$ tends to zero, namely:

$$\left|S^Z\right|^2 \leq \frac{N_A^2 + N_B^2}{k_B T}\left[\frac{\Omega}{p_0}\right] \tag{14}$$

where $\Omega$ is given by (12a), it means, by the equation constructed of the correlation functions strongly dependent on the structure of the sublattices appearing in the graphene as a basic element responsible for the existence of the requested structure. This result is directly connected with the linear dispersion characteristic for the graphene. Taking into account the nature of the graphene structure let us analyze the conditions for the existence of the magnetization from the physical point of view. The analysis is based on the formal result which can be formulated as follows: the magnetic order is expected in the two-dimensional geometry of the graphene structure even when the temperature is different from zero. The electron band configuration contains the linear term with respect to the wave vector whose appearance is determined by a non vanishing coefficient $\Omega$ given by (12a) in the dispersion law satisfied for a sample. The result remains in agreement with the calculations of Mermin-Wagner's theorem whose electronic density of the magnetic electrons band remains responsible for the magnetic behavior of a graphene construction of two interacting sublayers whose coupling plays an essential role in generation of magnetic field inside the graphene sheet.

According to (12a) we can see that the coefficient $\Omega$ is constructed by means of the correlation functions between the electrons belonging to two different structural sublayers. We can easily transform the considered correlation functions to the case when they are discussed in the diagonal representation and, due to this fact, can be interpreted as the occupation numbers which refer to the interacting pairs of electrons belonging to the same sublayer and having parallelly oriented spins. The electronic spectrum is determined by means of the linear coefficient which depends on the distribution of the electron occupation numbers in the lattice sites. The case when all the sites are occupied homogeneously in analogy to the classical situation the present model is reduced to its classical description. The magnetic order for $T > 0K$ vanishes. However, when the occupation numbers fluctuate from one lattice site to another we can see from the relation (12a) that the coefficient of the linear term in the electronic spectrum does not vanish and its appearance is responsible for the magnetic order existence above 0K. From the physical point of view the reason for which the graphene structure can be observed is connected with the inhomogeneous distribution of the electrons at the two dimensional layer. This inhomogeneity can have its explanation in the correlation effects. From the formal point of view the Mermin-Wagner's theorem works and the structure of graphene is given by the appearance of correlations still in two-dimensional space. The result of the present letter shows then the classical conclusion which is a confirmation of the Landau-Peiers statement that the two-dimensional crystal does not occur in the ideal conditions. Its stability depends on the fluctuating force constraints, or, in a more primitive form, in the case of anisotropic correlations in the two sublayers structure. In this way the diversity between the classical approach and new, recent result can be understood. The physical background allows us to construct the reality.

We have shown rigorously that the sufficient condition for the appearance of magnetization in the graphene single layer is connected with inhomogeneous distribution of spins between its two sublattices. This condition can be realized in two ways. First, in the ideal 2D graphene structure by stochastic distribution of electron correlations in two sublattices and next by the existence of a finite sublattice imbalance in the number of atoms belonging to each sublattice (zigzag, armchair edges [13, 14]) or by the missing atoms from any sublattice in an otherwise perfect system (vacancy [15, 16]).

From the physical point of view we can see that the homogenous distribution is not sufficient in order to create the magnetic field, the coefficient $\Omega$ is then tending to zero, which means that the magnetization in the temperature above $T = 0\text{K}$ is vanishing. Thus, in order to satisfy the Mermin-Wagner's condition containing the geometry, the two-dimensional graphene layer should be inhomogeneous with respect to the distribution of two sublayers sites. Thus, forming the appropriate superstructures in a graphene matrix can be a promising technique for engineering nanomaterials with desired magnetic properties [18]. Furthermore, the distribution of electrons, physically being a consequence of electronic fluctuations, seems to be one of the variants responsible for the existence of ferromagnetic order inside the graphene structure.